\documentclass{aa}

\usepackage{graphicx}
\usepackage{txfonts}
\usepackage{subcaption}
\usepackage{natbib}
\usepackage[group-minimum-digits=4]{siunitx}
\usepackage{gensymb}

\begin{document}

   \title{MRI-driven $\alpha\Omega$ dynamo at high Pm numbers}

   \author{Alexis Reboul-Salze
          \inst{1}, Loren E. Held\inst{2}, Kenta Kiuchi \inst{1,3}}

   \institute{\inst{1} Max Planck Institute for Gravitational Physics (Albert Einstein Institute), D-14476 Potsdam, Germany\\
    \email{alexis.reboul-salze@aei.mpg.de}  \\\inst{2} Department of Applied Mathematics and Theoretical Physics, Centre for Mathematical Sciences, University of Cambridge, Wilberforce Road, Cambridge CB3 0WA, United Kingdom\\
    \inst{3} Center for Gravitational Physics and Quantum Information, Yukawa Institute for Theoretical Physics, Kyoto University, Kyoto 606-8502, Japan
             }

   \date{Received ??; ??}

 
  \abstract
   {To power gamma-ray bursts and other high-energy events,  large-scale magnetic fields are required to extract rotational energy from compact objects such as black holes and neutron stars. The magnetorotational instability (MRI) is a key mechanism for angular momentum transport and large-scale magnetic field amplification. Recent work has begun to address the regime of high magnetic Prandtl number \(\mathrm{Pm}\), the ratio of viscosity to resistivity, in which angular momentum and magnetic energy increase with $\mathrm{Pm}$. This regime reveals unique dynamics of small-scale turbulence in disk mid-planes and buoyancy instabilities in the atmosphere. }
   {This study aims to build on these findings, focusing on the MRI-driven $\alpha\Omega$ dynamo in stratified simulations to understand magnetic field generation in the high-\(\mathrm{Pm}\) regime.}
   {We analyze data taken from stratified shearing box simulations both in the regime of magnetic Prandtl number of order unity, and also in the high $\mathrm{Pm}$ regime employing new techniques to compute the dynamo coefficients.}
   {We find that the mean-magnetic field evolution can be described by an $\alpha\Omega$ dynamo, even in the high-Pm regime. The mean magnetic field as well as the dynamo coefficients increase with Pm.
   This leads to a shorter dynamo period and a faster growth rate. We also find that the off-diagonal coefficients have an impact on the propagation of the magnetic field in the dynamo region. }
   {Overall, the magnetic field amplification found in global simulations should be increased by at least a factor of $~5$, which could lead to more powerful jets and stronger winds from astrophysical disks in the high-Pm regime.}

   \titlerunning{MRI-driven $\alpha\Omega$ dynamo at high Pm numbers}
   \authorrunning{A. Reboul-Salze, L. Held and K. Kiuchi}
   \maketitle
%

\section{Introduction}

After the merger of two neutron stars, the remnant object significantly impacts the electromagnetic counterparts of the event \citep{2017AbbottGRBandGW,2017GW170817Time,2017GoldsteinGRB170817,2017SavchenkoIntegralGRB170817,2018MooleySuperluminalGw170817}. 
Key factors include powerful neutrino luminosities, strong magnetic fields ($\geq 10^{15}$ G), and rotational energy reservoirs on the order of $10^{53}$ erg. These elements shape post-merger phenomena such as kilonovae, gamma-ray bursts, and relativistic jets \citep{2008MetzgerSGRB,2012BucciantiniSGRB,2013GiacomazzoSMNS,2013GaoBNSAfterglow,2014MetzgerSMNSemission,2014GompertzSGRB,2022SarinKilonovaSMNS}. Magnetic fields, in particular, play a critical role by extracting rotational energy and driving relativistic outflows from the merger remnant \citep{2020MostaMagnetarKilo, 2023CombiJetsfromBNS,2024Kiuchi,2024AguileraMiretDelayedJet,2024MusolinoHMNSJetlaunched}. 
In the case where the remnant collapses into a black hole, a large-scale magnetic field in the disk surrounding the black hole can launch relativistic jets that lead to the observed gamma-ray bursts \citep{2019FernandezGRMHDsim,2019ChristieBNSdisk,2022GottliebJetEjectaBNS,2023GottliebSecondsLongBHNSmerger,2023GottliebUnifiedPicture,2024HayashiBNSBHcase}. 

In addition to mergers, strong large-scale magnetic fields are critical in explaining extreme stellar explosions such as hypernovae and superluminous supernovae. Hypernovae are characterized by kinetic energies ten times higher than standard supernovae \citep{2011Drout}, while superluminous supernovae exhibit luminosities hundreds of times greater \citep{2013NichollSLSN,2018MargalitSLSN}. Hypernovae (broadline Type Ic supernovae) and superluminous supernovae are rare, representing $0.1\%$ and $1\%$ of all supernovae, respectively.
In both cases, a rapidly rotating protoneutron star (PNS) with a strong magnetic field can act as a central engine, efficiently extracting rotational energy to drive jets and magnetorotational explosions \citep[e.g.,][]{2006MoiseenkoMHDjet,2012WintelerHyper,2014MostaHyper,2018Obergaulinger,2020Bugli,2020KurodaMRexplosion3D}. Hypernovae can also be explained by the collapsar scenario, where the formation of a black hole and disk system can lead to relativistic jets \citep{1993WoosleyCollapsar,2009KomissarovCollapsar,2010JaniukCollapsar,2011TchekhovskoyMADstateBH,2021AloyCollapsars,2022Gottlieb,2023FujibayashiCollapsars}. 
Both central engines are associated with long and ultra-long gamma-ray bursts \citep{1992DuncanLGRB,2011MetzgerLGRB,2018MetzgerLGRBs}.

In these diverse astrophysical contexts, the magnetorotational instability (MRI) emerges as a promising mechanism for amplifying magnetic fields. The MRI operates efficiently even with weak seed fields and does not require a mean external field to sustain itself, as demonstrated in both unstratified \citep{1996HawleyMRINoFlux,2007FromangMRIB,2017RiolsMRIchimeras,2020MamatsashviliMRI} and vertically stratified simulations \citep{1995BrandenburgMRIdynamo,2010DavisButterfly,2010Shi,2015GresselMRImeanfield}
. This configuration, known as 'zero-net-flux'(ZNF) MRI, allows the exploration of MRI-driven turbulence independent of the geometry and strength of the central object's magnetic field or any external magnetic field threading the disk.

Explicit dissipation coefficients, including viscosity $\nu$ and resistivity $\eta$, are crucial for resolving the dynamics of MRI turbulence, particularly in stratified systems. In the case of ZNF configurations, ideal MHD simulations often suffer from convergence issues, where turbulent transport decreases with increasing resolution \citep{2007PessahMRI,2007FromangMRIB}. Including explicit dissipation restores convergence and reveals interesting and new physics. The ratio of the dissipation mechanisms, the magnetic Prandtl number $\mathrm{Pm}\equiv \nu/\eta$, is a key parameter for the ZNF MRI:
In the low magnetic Prandtl number (\(\mathrm{Pm}\)) regime (\(\mathrm{Pm} \leq 1\)), ZNF MRI struggles to sustain itself \citep{2015RiolsMRISustained}, while at high \(\mathrm{Pm}\), turbulent intensity exhibits power-law scaling \citep{2022GuiletHighPm,2022HeldMRIHighPm}. 
Several astrophysical objects that lead to high energy events are expected to be in the high \(\mathrm{Pm}\) regime such as the interiors of protoneutron stars \citep{2015GuiletVisc}, remnant neutron stars after binary neutron merger \citep{2017GuiletMergers} and  merger disks \citep{2008RossiPmdisk}, and the inner regions of x-ray binaries and AGN \citep{2008BalbusPmAGNdisk}. It is therefore crucial to study the large-scale magnetic field amplification in the high-Pm regime for these objects.

Stratified simulations reveal that the MRI is capable of generating large-scale magnetic fields through a dynamo mechanism, often interpreted as an $\alpha\Omega$ dynamo. This dynamo mechanism has been studied in shearing boxes \citep{2015GresselMRImeanfield,2016Shi}, spherical simulations of protoneutron stars \citep{2021ReboulSalze,2022ReboulSalze}, and global simulations of accretion disks in black hole-disk systems or binary neutron star mergers \citep{2024Kiuchi,2024HayashiBNSBHcase}. 
However, most of these studies have been limited to \(\mathrm{Pm}\) values near unity, with the highest explored being \(\mathrm{Pm} = 16\). Recent work by \citet{2024HeldMRIPm}, hereafter Paper I, extends these studies to \(\mathrm{Pm} \gg 1\) in stratified shearing box simulations, elucidating the interplay between small-scale MRI-driven turbulence in the mid-plane and large-scale Parker-unstable structures in the atmosphere.

In this study, we build on the findings of Paper I, focusing on the MRI-driven $\alpha\Omega$ dynamo in stratified simulations at high \(\mathrm{Pm}\). By employing new diagnostic tools, we aim to deepen our understanding of how large magnetic Prandtl numbers influence dynamo action with a focus on the large-scale magnetic field generation for neutron stars, accretion disks, and related astrophysical systems.
The paper is organized as follows: In section \ref{sec:model}, we describe the numerical setup and the methods used to compute the dynamo coefficients. In section \ref{sec:Results} we analyze shearing-box simulations in the high-Pm regime. We then compare these results from local simulations to a recent high-resolution global GRMHD simulation in section \ref{sec:Comp}. 
Lastly, we discuss the validity of our assumptions in section \ref{sec:Disc}. We conclude in section \ref{sec:Conclusion}.

\section{Methods and numerical setup}
\label{sec:model}

\subsection{Shearing Box Simulations}

Our numerical approach utilizes the shearing box model, which approximates a small, rotating patch of an accretion disk using a Cartesian grid. This local framework allows for the detailed study of MRI-driven turbulence and dynamo action without global disk or star complexities. 

The governing equations for magnetohydrodynamics (MHD) in the shearing box approximation are:
\begin{align}
\frac{\partial \rho}{\partial t} + \nabla \cdot (\rho \mathbf{u}) &= 0, \\
\frac{\partial \mathbf{u}}{\partial t} + (\mathbf{u} \cdot \nabla)\mathbf{u} &= -\frac{\nabla P}{\rho} - 2\mathbf{\Omega} \times \mathbf{u} + \mathbf{g}_{\text{eff}} + \frac{1}{\mu_0 \rho} (\mathbf{J} \times \mathbf{B}) + \nabla \cdot \mathbf{T} \\
\frac{\partial \mathbf{B}}{\partial t} &= \nabla \times (\mathbf{u} \times \mathbf{B})+\eta \nabla^2 \mathbf{B},
\end{align}
where \(\mathbf{T}\) is the viscous stress tensor, $\mathbf{g}_{\text{eff}}$  the tidal acceleration, and \(\mathbf{J} = \nabla \times \mathbf{B}\) is the current density. 
The simulations are vertically stratified and the tidal acceleration $\mathbf{g}_{\rm eff} = 2 q \Omega_0^2 x\vec{e_x} -\Omega_0^2 z \vec{e_z}$,  where $q=- \frac{d \ln \Omega}{d \ln \ r}$ is the shear rate and $\Omega_0$ is the angular frequency at the radius about which the box is centered, embodies the effective gravitational potential. We use $q=1.5$, which corresponds to the Keplerian disk case, while in a hypermassive neutron star, the shear rate can be lower, ranging from $q \sim 1$ to $q \sim 1.5$.

We use the \textsc{PLUTO} code with explicit diffusion terms to control the viscosity \(\nu\) and the resistivity \(\eta\).  
We adopt the zero-net-flux (ZNF) configuration, initializing the magnetic field as \(\mathbf{B}_0 = B_0 \sin(2\pi x/L_x) \hat{z}\), with a plasma beta (\(\beta_0 \equiv 2\mu_0P_0/B_0^2\)) of 1000. We adopt an isothermal equation of state $P_0 = \rho_0 c_{s0}^2$, where $P_0$, $\rho_0$, and $c_{s0}$ are the initial mid-plane pressure, density, and isothermal sound-speed, respectively. The box dimensions are set to \(4H \times 4H \times 8H\), with \(128\) grid cells per scale height \(H=c_{s0}/\Omega_0 \) to ensure adequate resolution.

In this paper, we focus on the y-averaged quantities in order to facilitate comparison with the corresponding axisymmetric averages in global simulations. We compare data from two shearing-box simulations from Paper I 
at $\mathrm{Pm}=4$, representing the regime of $\mathrm{Pm} = O(1)$, and at $\mathrm{Pm}=90$, representing the high Pm regime. Further details of the turbulence and spectra can be found in Paper I. 
The snapshots used to compute the averaged quantities are output every $\Delta t_{\rm snap}= 0.1$ orbits. 

\subsection{Mean-Field Theory}

The evolution of the averaged magnetic field can be described using mean-field theory. The general mean-field theory has been developed by \citet{1978moffattfield} and \citet{1980KrauseMeanField} and has been widely used to study dynamos \citep{2018BrandenburgMeanFieldReview, 2019RinconDynTh}.
We use the mean-field concept to understand which processes dominate the generation of the mean magnetic field.
The basic idea of a mean-field dynamo is that a large-scale magnetic field is generated by small-scale turbulence. The velocity and magnetic fields are therefore decomposed into a mean and a small-scale component, which we represent using the following notation: $\vec{X}=\overline{\vec{X}}+\vec{X}'$. The definition of \emph{mean} here is the axisymmetric average operator noted $\overline{\cdot}$, which satisfies the Reynolds averaging rules.  The approach of mean-field theory is to expand the electromotive force (EMF) $\overline{\vec{\mathcal{E}}}$ in terms of the mean quantities ($\overline{\vec{u}}$ and $\overline{\vec{B}}$) and the statistical properties of the fluctuating quantities ($\vec{u}'$ and $\vec{B}'$). 
In mean-field theory, the mean magnetic field \(\overline{\mathbf{B}}\) evolves under:
\begin{equation}
\frac{\partial \overline{\mathbf{B}}}{\partial t} = \nabla \times (\mathbf{\overline{\mathcal{E}}} + \overline{\mathbf{B}} \times \overline{\mathbf{U}}) + \eta \nabla \times (\nabla \times \overline{\mathbf{B}}),
\end{equation}
where \(\overline{\mathbf{\mathcal{E}}} = \overline{ \mathbf{u'} \times \mathbf{b'}}\) 
represents the turbulent electromotive force (EMF), which encapsulates the small-scale feedback of velocity and magnetic field fluctuations. 
To depend only on the mean quantities, the \(\mathbf{\overline{\mathcal{E}}}\) is parameterized as:
\begin{equation}
\overline{\mathcal{E}}_i = \alpha_{ij} \overline{\mathbf{B}}_j + \beta_{ij} (\nabla \times \overline{\mathbf{B}})_j,
\end{equation}
where \(\alpha_{ij}\) and \(\beta_{ij}\) are the dynamo coefficients of tensors that do not depend on $\overline{\vec{B}}$.

In this subsection, we work in cylindrical polar coordinates $(s, \phi, z)$.
Previous studies showed that MRI-driven dynamos can be interpreted as so-called $\alpha\Omega$ dynamos \citep{2022ReboulSalze, 2024DhangMRI_dynamo,2024Kiuchi}.  The $\Omega$ effect corresponds to the shearing of the magnetic field by differential rotation. 
It generates toroidal magnetic fields from the poloidal field. With cylindrical differential rotation, the $\Omega$ effect reads
\begin{equation}
      \frac{\partial\overline{B}_\phi}{\partial t} = s \overline{B}_s \frac{d \Omega}{d s}
,\end{equation}
and it induces an anti-correlation between the radial field $\overline{B}_s$ in cylindrical coordinates and the toroidal magnetic field~$\overline{B}_\phi$. 

The $\alpha$ effect comes instead from the closure relation of the mean EMF, shown above. The effect of the diagonal $\alpha$ coefficients
can be described as the twisting of the mean magnetic field lines by turbulence, which then forms magnetic field loops that can generate a poloidal magnetic field from the toroidal magnetic field and vice versa.

In our case, the generation of the poloidal magnetic field by this effect can be seen as a correlation between the toroidal component of the EMF and the toroidal component of the magnetic field $\overline{B}_\phi$ in the form of ${\overline{\mathcal{E}}}_\phi = \alpha_{\phi \phi} \overline{B}_\phi$. 
The diagonal components of the $\beta$ tensor are in the direction of the mean current $\overline{\vec{J}} = \mu_0^{-1} \nabla \times \overline{\vec{B}}$, and their effect can be described as a turbulent diffusivity, which adds to the physical magnetic diffusivity $\eta$.

By definition of local shearing boxes in Cartesian coordinates, there is equivalence in terms of coordinates $(x,y,z) = (s-s_0, s_0(\phi-\phi_0-\Omega_0 t), z)$, where $s_0$ is the radius about which the box is centered. 

\subsection{Methods for Computing Dynamo Coefficients}

\begin{figure*}[ht]
    \centering
\begin{subfigure}[b]{0.49\textwidth}
\includegraphics[width=1.0\textwidth]{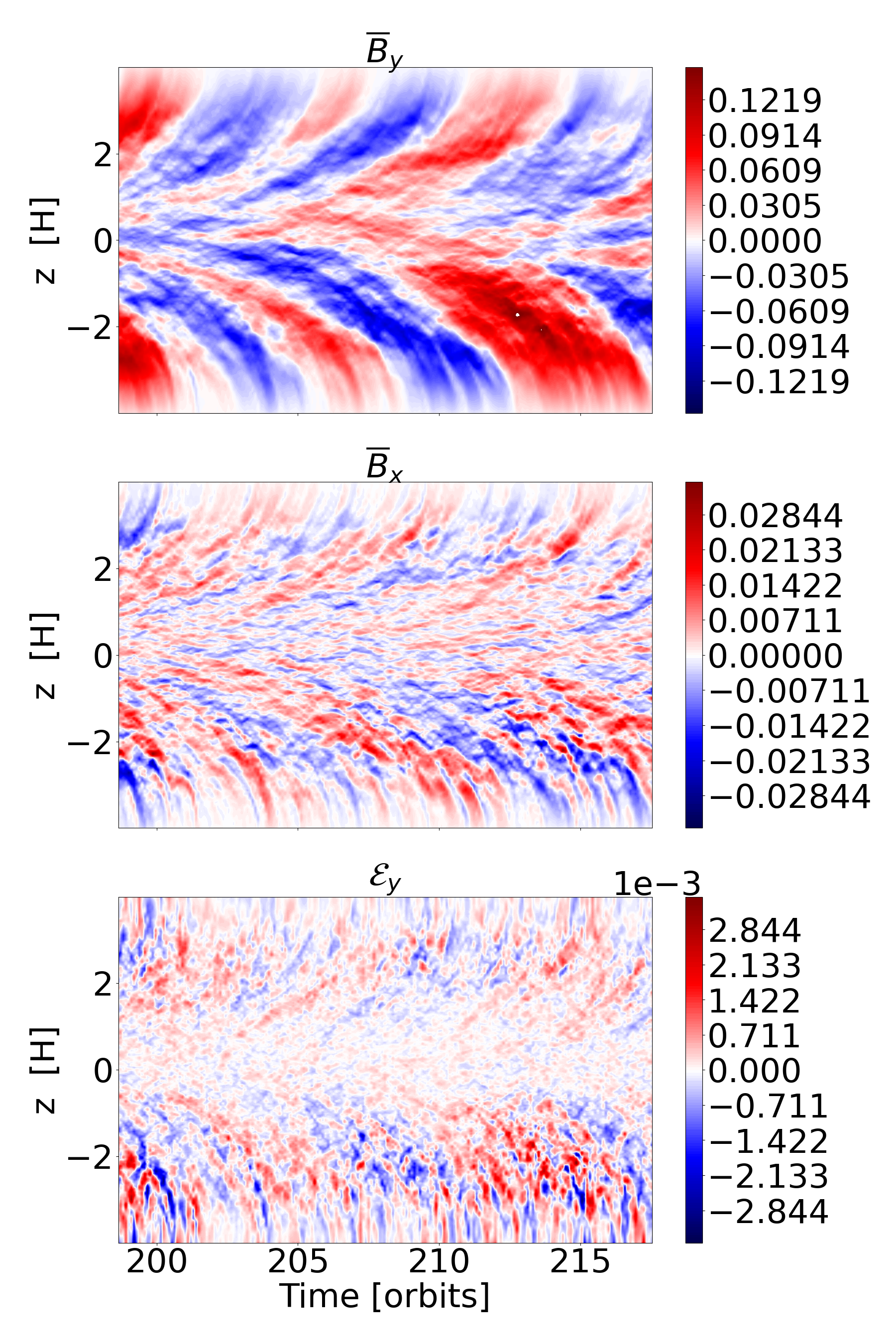}
     \caption{Pm=4}
\end{subfigure}
\begin{subfigure}[b]{0.49\textwidth}
\includegraphics[width=1.0\textwidth]{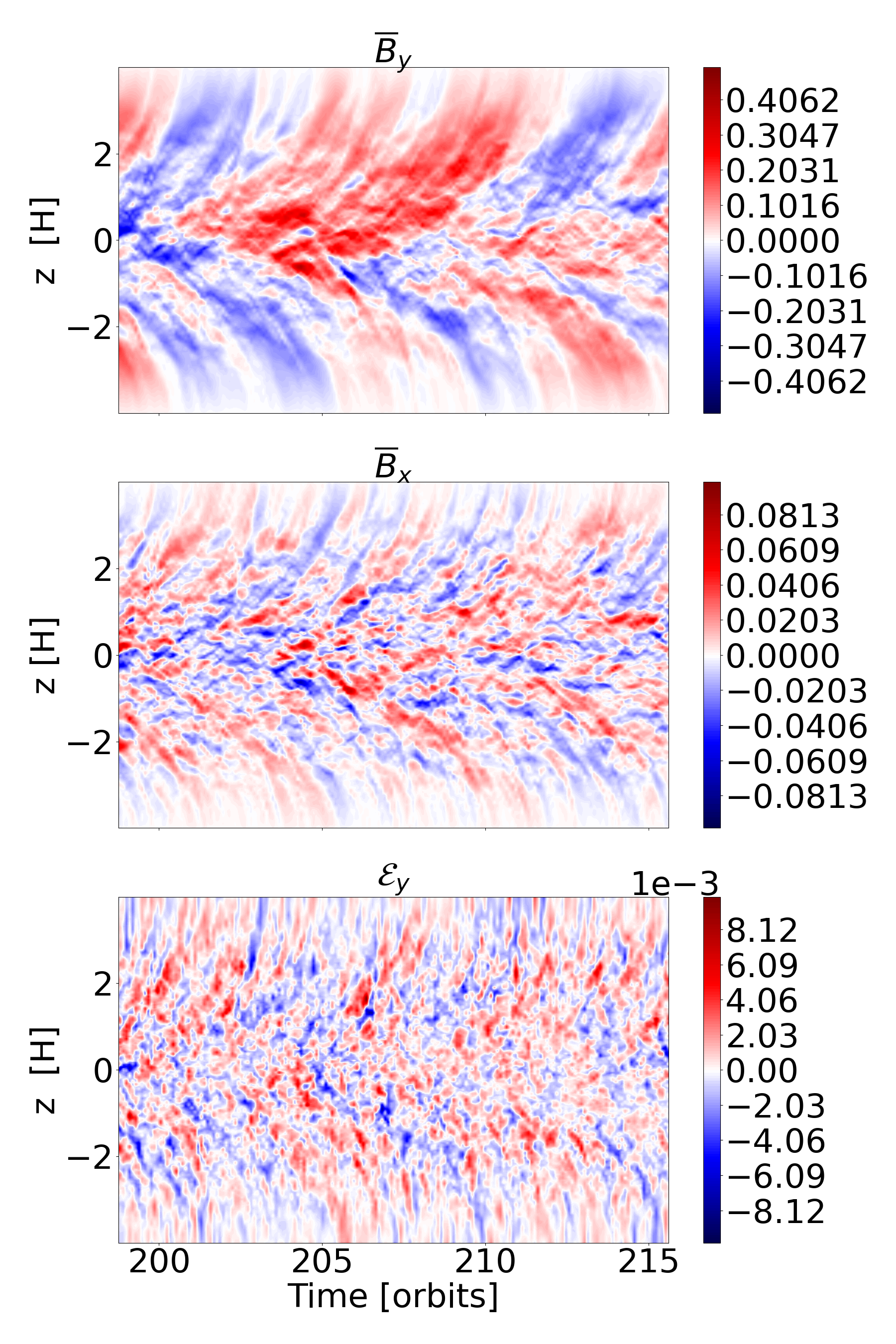}
    \caption{Pm=90}
\end{subfigure}
\caption{Butterfly diagrams of the toroidal magnetic field (top), the radial magnetic field (middle), and the toroidal component of the EMF $\overline{\mathcal{E}}_y$ (bottom) from simulations run with magnetic Prandtl numbers $\mathrm{Pm}=4$ (left) and $90$ (right), respectively.}
\label{fig:butterfly}
\end{figure*}

In this paper, we use and compare three different methods to compute the dynamo coefficients: the correlation method, SVD decomposition, and the IROS algorithm.
As a diagnostic for the correlations between the magnetic field and the electromotive force, we use the Pearson correlation coefficient between two quantities $X$ and $Y$ following:
\begin{equation}
    \mathcal{C}_P(X,Y) = \frac{\int_t (X - \langle X \rangle_t) dt \int_t (Y - \langle Y \rangle_t) dt }{\sqrt{(\int_t (X - \langle X \rangle_t)^2 dt)}\sqrt{(\int_t (Y - \langle Y \rangle_t)^2 dt)}}
    \label{eq:correlation}
,\end{equation}
where $\langle \cdot \rangle_t$ represents a time average.
As a first-order estimate for the dynamo coefficients of, for example, the $\alpha$ tensor, we use what we will call the correlation "method" with the formula
\begin{equation}
    \alpha_{ii} = \frac{\langle \overline{B_i}  \ \overline{\mathcal{E}_i}\rangle_t}{\langle  (\overline{ B_{i}})^2 \rangle_t}\,.
    \label{e:alpha}
\end{equation}
This estimate assumes that the EMF is only due to the $\alpha$ effect, which is a good approximation in the case of high correlation values. However, it tends to overestimate the coefficients as it assumes that the estimated coefficient is the only source of the electromotive force. 

In order to take into account the off-diagonal contributions, we compute the values of the alpha tensor coefficients in this study by using the singular value decomposition (SVD) method to perform the least-square fit of mean-current data and mean field \citep{2011Racine}.

To compute all the dynamo coefficients simultaneously, we use the Iterative Removal of Sources (IROS) method \citep{1992Hammersley}. This algorithm has been developed for radio astronomy and applied to isolate contributions to the turbulent EMF from mean magnetic fields and currents \citep{2024DhangMRI_dynamo,2024BendreIROS}. The key steps include:
\begin{enumerate}
    \item \textbf{Initialization:} Start with zero values for \(\alpha_{ij}\) and \(\beta_{ij}\).
    \item \textbf{Fitting:} At each spatial position, fit the EMF time series against \(\mathbf{B}\) and \(\mathbf{J}\) to extract first-order estimates.
    \item \textbf{Refinement:} Iteratively subtract a percentage of the best-fit contributions from the EMF and update coefficients until convergence.
\end{enumerate}
This method provides robust estimates even in the presence of noise and permits the inclusion of priors, enhancing its applicability in high-resolution simulations of MRI-driven dynamos.

\section{Results}
\label{sec:Results}

In this paper, we focus  on the dynamics of the axisymmetric quantities, which correspond to the y-averaged quantities in shearing boxes. Since shearing boxes are local models, we also averaged in the x direction in order to filter some of the noise due to turbulence.
It was found in Paper I  that the simulation could be divided in two regions where the dynamics are governed by the MRI dynamo or by the magnetic buoyancy, respectively the "MRI dynamo" zone and the "atmospheric zone". 
In the following, these two zones will be defined as $z \in [-2H,2H]$ for the dynamo zone and $z \in [-4H,-2H] \ \rm and \ [2H,4H]$ for the atmospheric zone.

\subsection{Butterfly diagrams}

We first show the butterfly diagrams (Figure \ref{fig:butterfly}) of the three important quantities to check whether it is an $\alpha\Omega$ dynamo: $\bar{B}_y$, $\bar{B}_x$ for the $\Omega$ effect and  $\bar{B}_y$, $\overline{\mathcal{E}}_y$ for the $\alpha$ effect.
The averaged magnetic field is around 3 times stronger at $\mathrm{Pm}=90$ than at $\mathrm{Pm}=4$. This value is expected since, in Paper I, the ratio of magnetic energy in the MRI dynamo zone is $E_{\rm mag,Pm=90}/E_{\rm mag,Pm=4} \approx 7.8$. The factor for the magnetic field between the two simulations is therefore $\sim 2.8$. 
Note that this is slightly lower than what the magnetic energy scaling $E_{\rm mag} \propto \mathrm{Pm}^{0.744}$ would predict because the plateau at high-Pm starts to set in.
This shows that it is not only the turbulent field that is further amplified with Pm but also the largest scale of the simulation, as was found by \citet{2022GuiletHighPm}. 
In both simulations, the butterfly diagram of the averaged toroidal magnetic field $\bar{B}_y$ shows some cycles. In the case of less turbulent $\mathrm{Pm}=4$ simulation, the cycles are very regular while at $\mathrm{Pm}=90$ they are more chaotic and the period is harder to decipher. $\overline{B}_y$ seems to be at larger scales for $\mathrm{Pm}=4$ than $\mathrm{Pm}=90$, while Paper I  found the turbulence to be at larger scales for $\mathrm{Pm}=90$. 
This may seem contradictory initially but it can be explained by the more intense turbulence at $\mathrm{Pm}=90$ which is more efficient at generating (as can be seen by comparing the colorbars at $\mathrm{Pm}=4$ and $\mathrm{Pm}=90$ in Figure \ref{fig:butterfly}) and therefore at destroying the coherent magnetic fields $\overline{B}_{x}$ and $\overline{B}_{y}$ (as can be seen by the more turbulent cycles at $\mathrm{Pm}=90$). 

Once the coherent magnetic field reaches the atmosphere beyond $|z|\ge 2H$, the magnetic field behaves similarly and is as coherent and on a similar scale for both $\mathrm{Pm}$.
The result of the turbulence at larger scales is recovered for the other quantities $\overline{B}_x$ and $\overline{\mathcal{E}}_y$.
Another difference that can be seen in the $\overline{B}_y$ butterfly diagrams is the propagation velocity towards the atmosphere: at $\mathrm{Pm}=4$, the propagation is slower than at $\mathrm{Pm}=90$.
However, once in the atmosphere, the magnetic field seems to propagate at very similar velocities. 
This seems to be coherent with the definition of the atmosphere, where dynamics are dominated by the buoyancy instabilities rather than by the MRI-driven dynamos.

For the $\mathrm{Pm}=4$ simulation, the anti-correlations between $\overline{B}_y$ and $\overline{B}_x$ can be seen directly in the butterfly diagram. This correlation is harder to see at $\mathrm{Pm}=90$ as the radial magnetic field is more turbulent. Nevertheless, it is possible to see the correlation closer to the atmosphere, where the averaged magnetic field is less turbulent. 
The periodic cycle is also difficult to see in the $\overline{B}_x$ diagram for the $\mathrm{Pm}=90$ case. For the EMF, the difference is striking between the two simulations: in the midplane, the y-component of the EMF is weaker than in the atmosphere for $\mathrm{Pm}=4$, while it is of similar amplitude for $\mathrm{Pm}=90$
, which further shows the increased mean-field dynamo action with Pm. In addition, the cycle can clearly be seen for both simulations and they seem to be correlated with $\overline{B}_y$ diagrams. The MRI-driven mean-field dynamo can be qualitatively interpreted as an $\alpha\Omega$ dynamo.

Lastly, from these diagrams, we estimate the simulation period by calculating time Fourier transforms of $B_x$ at vertical positions  $z \in [-2H,2H]$, and averaging the frequencies with the highest magnitude. We choose to use $\overline{B}_x$ as the $\alpha$ effect governs the generation of $\overline{B}_x$. The frequencies are shown in Table \ref{tab:dyn_period}. The period is slightly shorter in the case of $\mathrm{Pm}=90$, which would be consistent with the stronger dynamo action. Note that there is some variability in the periods due to the short time where high-cadence snapshots are available but all frequencies at $\mathrm{Pm}=90$ were systematically higher than at $\mathrm{Pm}=4$.
Indeed, the period of an $\alpha\Omega$ dynamo is given by \citep{2015GresselMRImeanfield}
\begin{equation}
    P_{\alpha\Omega} = \frac{2\pi}{\omega_{\alpha \Omega}} = \frac{2\pi}{\left|\frac{1}{2} \alpha_{yy}\frac{d \Omega}{d \ln \ s} k_z\right|^{1/2}},
    \label{eq:dyn_period}
\end{equation}
where $k_z$ is the vertical wavenumber with as $k_z= \frac{2\pi}{H}$.
In order to check quantitatively whether it is correct, we need to compute the dynamo coefficients.

\begin{table}[]
    \renewcommand{\arraystretch}{1.2}
    \caption{Periods of the different mean-field patterns above or below the midplane ($z=0$) 
    for $\mathrm{Pm}=4$ and $\mathrm{Pm}=90$. The period is estimated by taking the FFT of $\overline{B_x}$ at $|z|=2 H$. $P_{\alpha\Omega}$ is the theoretical period taken with the maximum value of $\alpha_{yy}$ above or below the midplane.} 
    \label{tab:dyn_period}
    \centering
    \begin{tabular}{lcccc}
    \hline\hline
     $\mathrm{Pm}$ & $P_\mathrm{above}$ &$P_\mathrm{below}$ & $P_{\alpha\Omega,N}$ & $P_{\alpha\Omega,S}$  \\
      - & [orbits] & [orbits] & [orbits] & [orbits] \\
    \hline
    4 & 8.33 & 5.47& 6.2 & 6.23\\
    90 & 5.55 & 3.73 & 4.9 & 4.03 \\ 
    \hline \hline
    \end{tabular}
\end{table}

\subsection{$\alpha$ effect and dynamo period}

\begin{figure}[ht]
\includegraphics[width=0.45\textwidth]{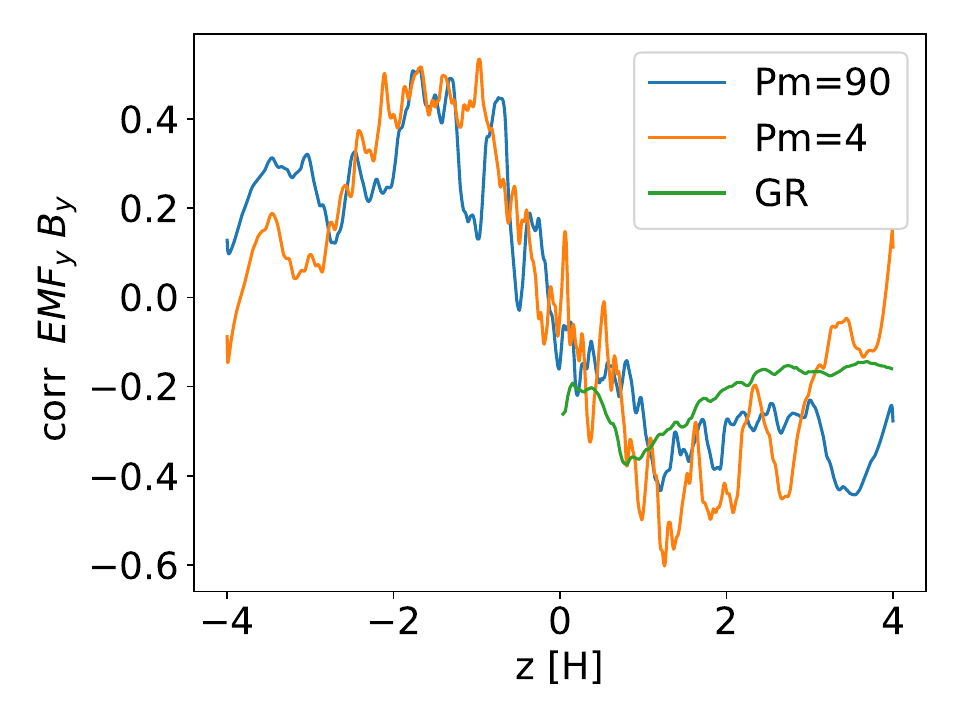}
\includegraphics[width=0.45\textwidth]{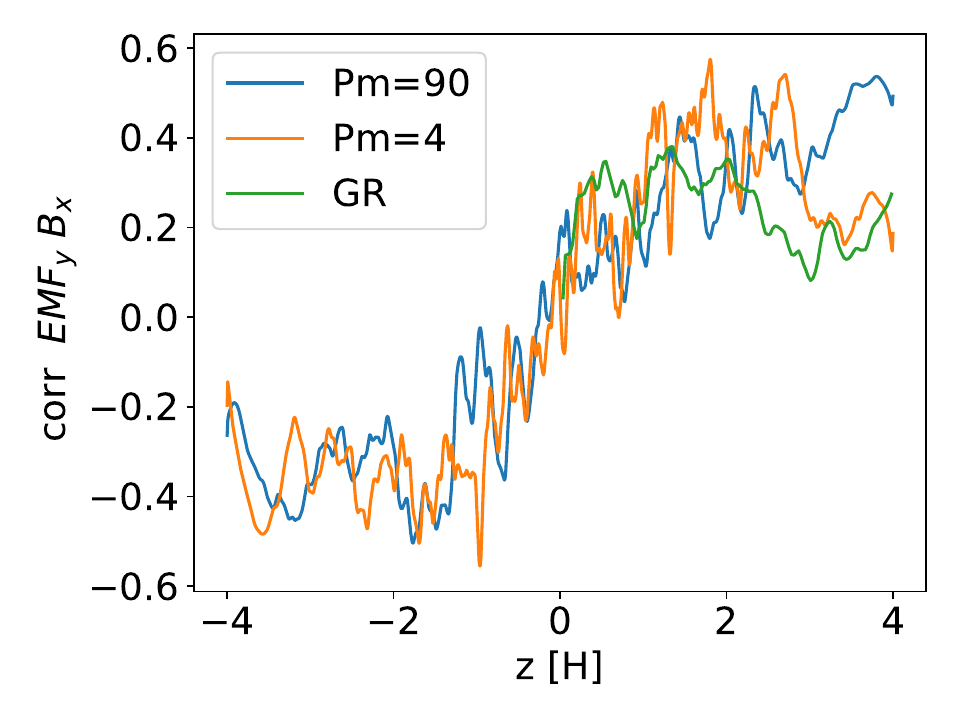}
\includegraphics[width=0.45\textwidth]{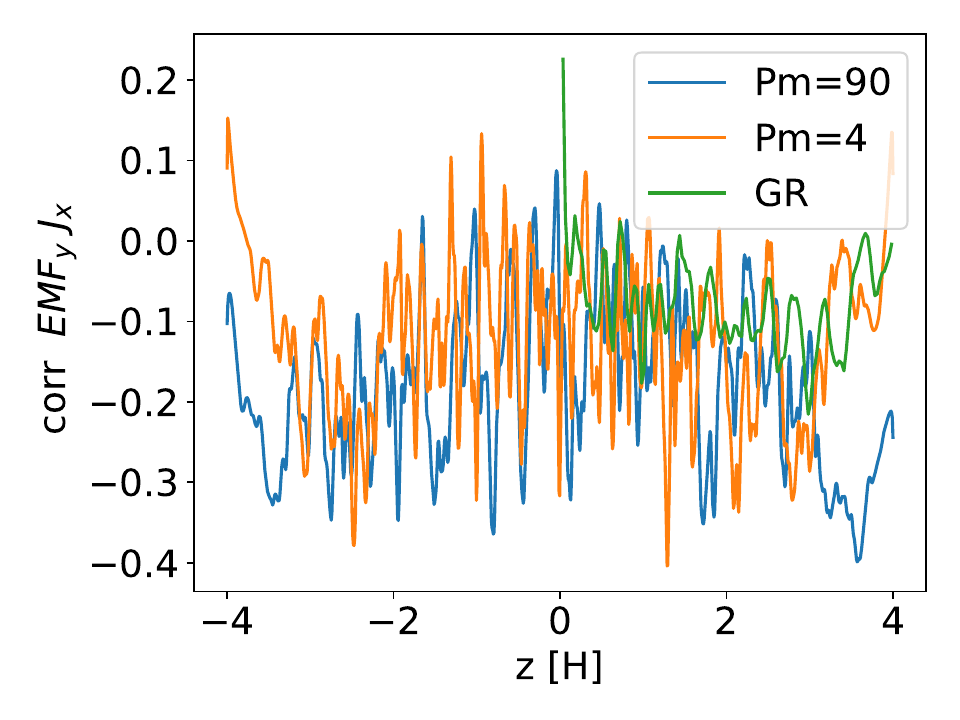}
\caption{Pearson correlations between $\overline{\mathcal{E}}_y$ and $\bar{B}_y, \text{(top)}\ \bar{B}_x$ (middle) and $\bar{J}_x$ (Bottom). The green curve corresponds to the GRMHD global simulation described in \citet{2024Kiuchi}, where the vertical length was rescaled by $H_{GRMHD}=7$ km.
}
\label{fig:corr_comp}
\end{figure}

To confirm the effect of $\mathrm{Pm}$ on the $\alpha\Omega$ dynamo, we compute the correlations to check whether the diagonal alpha effect is the dominant contribution to the toroidal $\overline{\mathcal{E}}_y$ (Figure \ref{fig:corr_comp}). It shows that there is a strong correlation between $\overline{B}_y$ and $\overline{\mathcal{E}}_y$ for $|z| \leq 2.5H$, which decreases in the atmosphere. 
For $|z|\geq 2.5H$, the correlation between $\overline{B}_y$ and $\overline{\mathcal{E}}_y$ becomes non-dominant compared to the correlation between $\overline{B}_x$ and $\overline{\mathcal{E}}_y$. These results confirm that the dynamics of the atmosphere and the dynamo region of the shearing box are quite different. 
We also find that the correlation between the radial resistivity $\overline{J}_x$ and the toroidal $\overline{\mathcal{E}}_y$ is not important in the dynamo region, which means that the turbulent resistivity can be neglected for the generation of the poloidal field from the toroidal field. 
In the correlations, the differences are quite small between $\mathrm{Pm}=4$ and $\mathrm{Pm}=90$. Due to the less turbulent dynamo, the correlations are slightly higher for $\mathrm{Pm}=4$ but remain overall similar.

\begin{figure}[ht]
\includegraphics[width=0.45\textwidth]{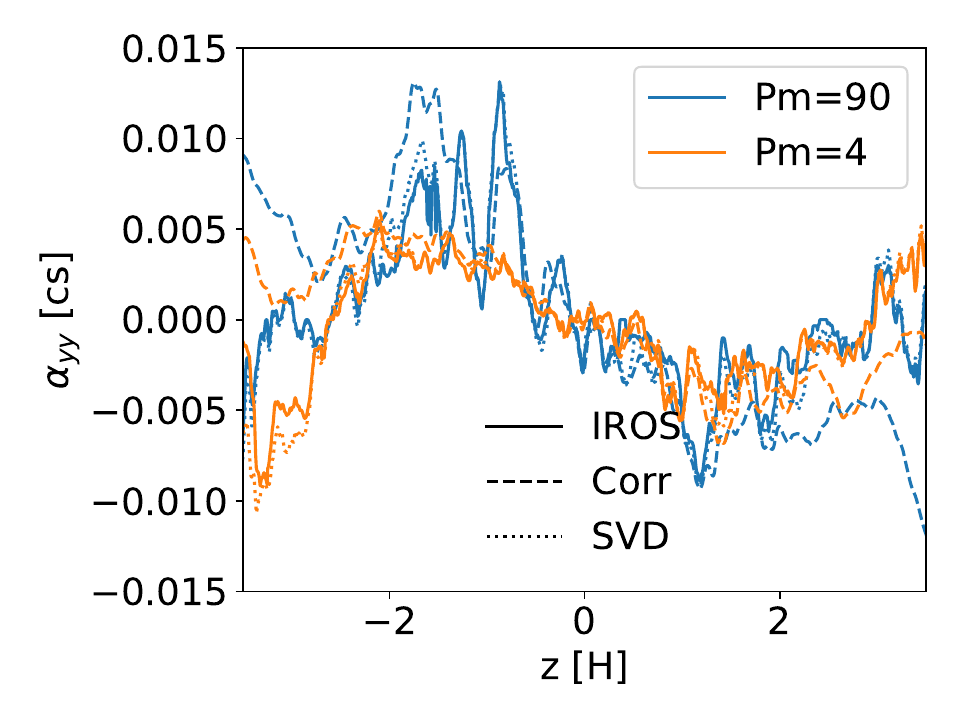}
\includegraphics[width=0.45\textwidth]{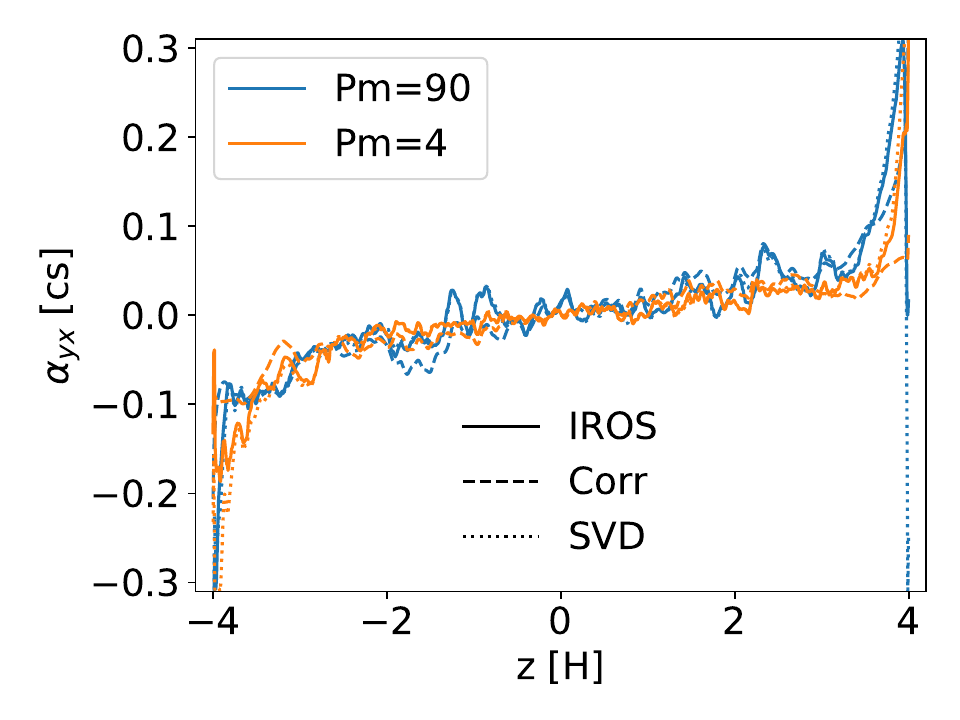}
\caption{$\alpha$ dynamo coefficients as a function of  vertical coordinate $z$. The solid, dashed, and dotted curves are calculated using the IROS, correlation (Corr), and SVD methods, respectively. Orange curves: $\mathrm{Pm}=90$. Blue curves: $\mathrm{Pm}=4$.} 

\label{fig:alpha_Pmcomp}
\end{figure}

The correlations hint that the dynamo cycle could be interpreted as an $\alpha\Omega$ dynamo and that turbulent resistivity could be neglected to first-order for the EMF. To see whether we recover the difference in the dynamo period between $\mathrm{Pm}=4$ and $\mathrm{Pm}=90$ in the dynamo coefficients, we compute the $\alpha$ dynamo coefficients with the IROS method, the correlation methods, and the SVD method (Figure \ref{fig:alpha_Pmcomp}). With the IROS and correlation methods, the maximum value of the diagonal $\alpha_{yy}$ effect is stronger by a factor $\approx 3$ in the case of $\mathrm{Pm}=90$. The more chaotic aspect of the turbulence in $\mathrm{Pm}=90$ can also be seen with the stronger oscillations of the coefficient along the vertical axis.
The coefficient values for $\alpha_{yy}$ are stronger with the correlation methods in the region where the correlations are weaker as expected since the EMF is considered to be explained by only one dynamo coefficient at a time. The SVD method gives intermediate results between the IROS method and correlation methods. 
This corresponds to the atmosphere region for the diagonal $\alpha_{yy}$. 
For $\alpha_{yx}$, the main difference is close to the atmosphere boundaries, where SVD and IROS methods predict a stronger increase than the correlation method. Otherwise, all the methods roughly agree, except around $|z| \sim 2H$  where $\alpha_{yx}$ is slightly stronger with the correlation method.

Using the maximum value of $\alpha_{yy}$ in the dynamo region, we compute the dynamo period with the theoretical formula from mean-field theory (equation \ref{eq:dyn_period}).
The theoretical periods give a good agreement with what can be found in the butterfly diagrams as shown in Table~\ref{tab:dyn_period}. The difference in frequencies between above or below the midplane was not recovered for the $\mathrm{Pm}=4$ simulation. It might be due to the stochasticity of the turbulence over short times.

These results show that the large-scale dynamo can be interpreted as an $\alpha\Omega$ dynamo and it gets stronger with increasing $\mathrm{Pm}$.

\subsection{Off-diagonal $\alpha$ effect and propagation of the magnetic field}

Another finding from the correlations is that the off-diagonal coefficient $\alpha_{yx}$ is important in the atmosphere region ($|z|>2 H$). By using the value of the coefficient that is increasing away from the midplane (Bottom panel of Figure \ref{fig:alpha_Pmcomp}), we can estimate the height at which the contribution of $\alpha_{yx} \overline{B}_x$ to the EMF becomes more important than the diagonal coefficient $\alpha_{yy} \overline{B}_y$. It gives a $|z_{\rm off-diag}| \approx 2-2.5H$, which corresponds roughly to the transition from the dynamo region to the atmosphere region. This explains why the dynamo period is well explained by its theoretical value depending only on $\alpha_{yy}$, even with the correlation between $\overline{B}_x$ and $\overline{\mathcal{E}}_y$.
To try to understand the physical impact of this off-diagonal coefficient, we compute the vertical turbulent pumping coefficient 
\begin{equation}
    \gamma_z = \frac{1}{2}\left(\alpha_{xy}-\alpha_{yx}\right).
\end{equation}
In the closure relation of the EMF, the turbulent pumping $\vec{\gamma}$ is the antisymmetric component of the general $\alpha$ tensor and can therefore be written as 
\begin{equation}
    \overline{\mathcal{E}}= \vec{\gamma} \times \overline{B} + [...].
\end{equation}
The turbulent pumping acts then exactly as a velocity on the averaged magnetic field.

\begin{figure}[ht]
\includegraphics[width=0.45\textwidth]{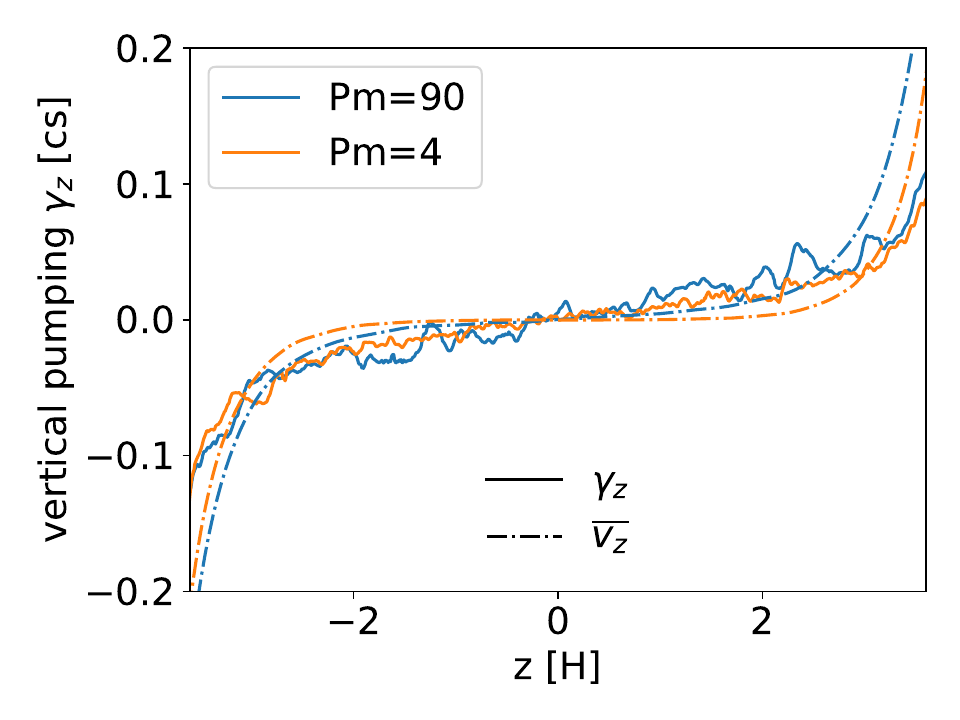}
\caption{Turbulent pumping coefficient $\gamma_z$ calculated using the IROS method (solid lines), and averaged vertical velocity $\bar{v}_z$ (dashed lines), as a function of of the vertical coordinate $z$.}
\label{fig:Turbulent_pump_comp}
\end{figure}

We compare the averaged vertical velocity $\overline{v}_z$ and $\gamma_z$ in Figure \ref{fig:Turbulent_pump_comp}. 
Once again, the difference between the dynamo region and the atmosphere region is striking. In the atmosphere region, the propagation of the magnetic field is dominated by the average velocity $\overline{v}$, which is due to the buoyancy instabilities and therefore linked to the mean magnetic field strength. This explains why the vertical velocity is faster for $\mathrm{Pm}=90$.
In the dynamo region, the vertical propagation is dominated by the turbulent pumping due to the small-scale turbulence.  It is also stronger in the high-$\mathrm{Pm}$ case, which explains why the magnetic field in the butterfly diagram is moving faster towards the atmosphere.
This shows the importance of the off-diagonal coefficient for the disk or neutron star dynamics, as a faster propagation of stronger magnetic fields towards the atmosphere could increase the efficiency of the disk or neutron star winds.

\subsection{Turbulent angular momentum transport}

\begin{figure*}[!h]
\centering
\includegraphics[width=0.49\textwidth]{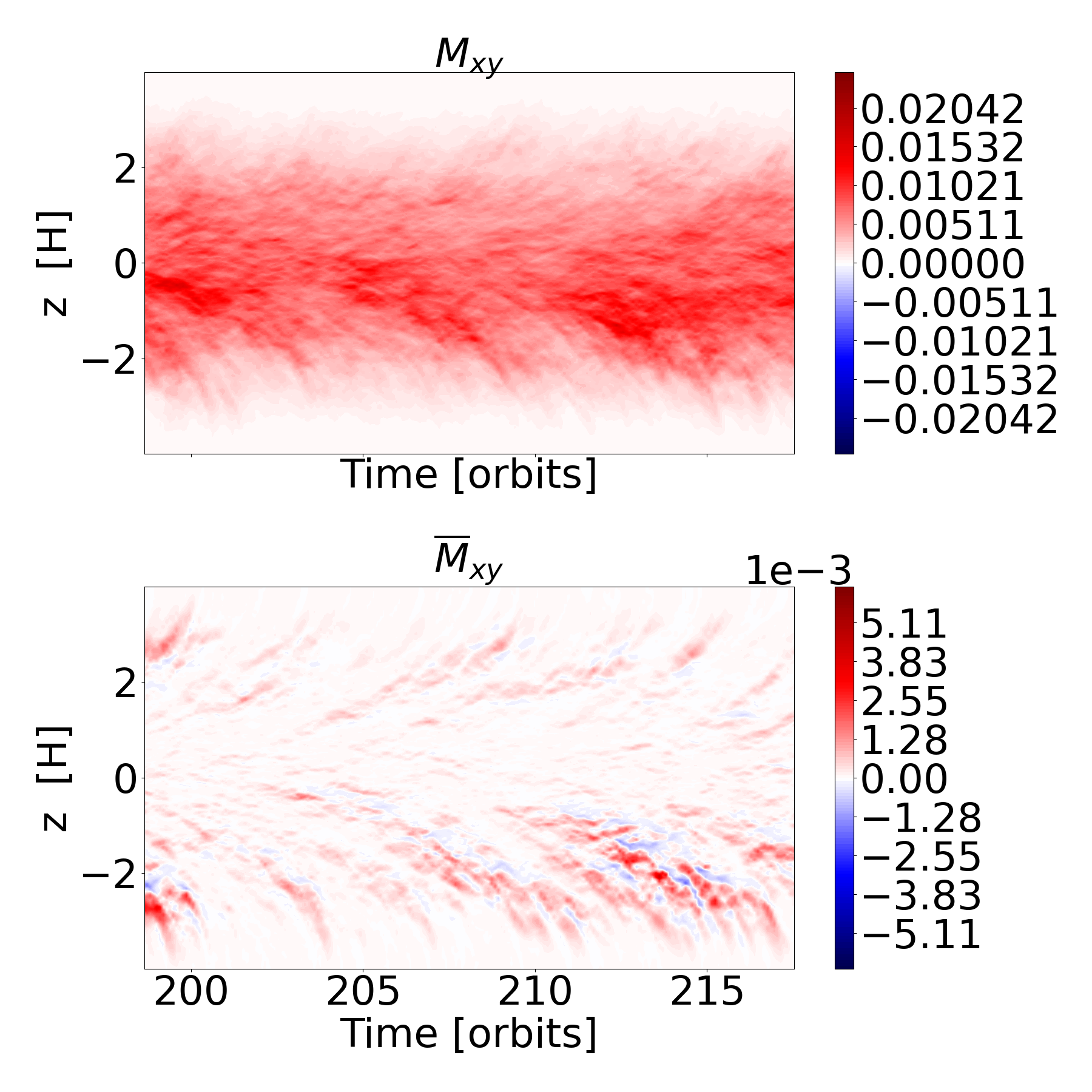}
\includegraphics[width=0.49\textwidth]{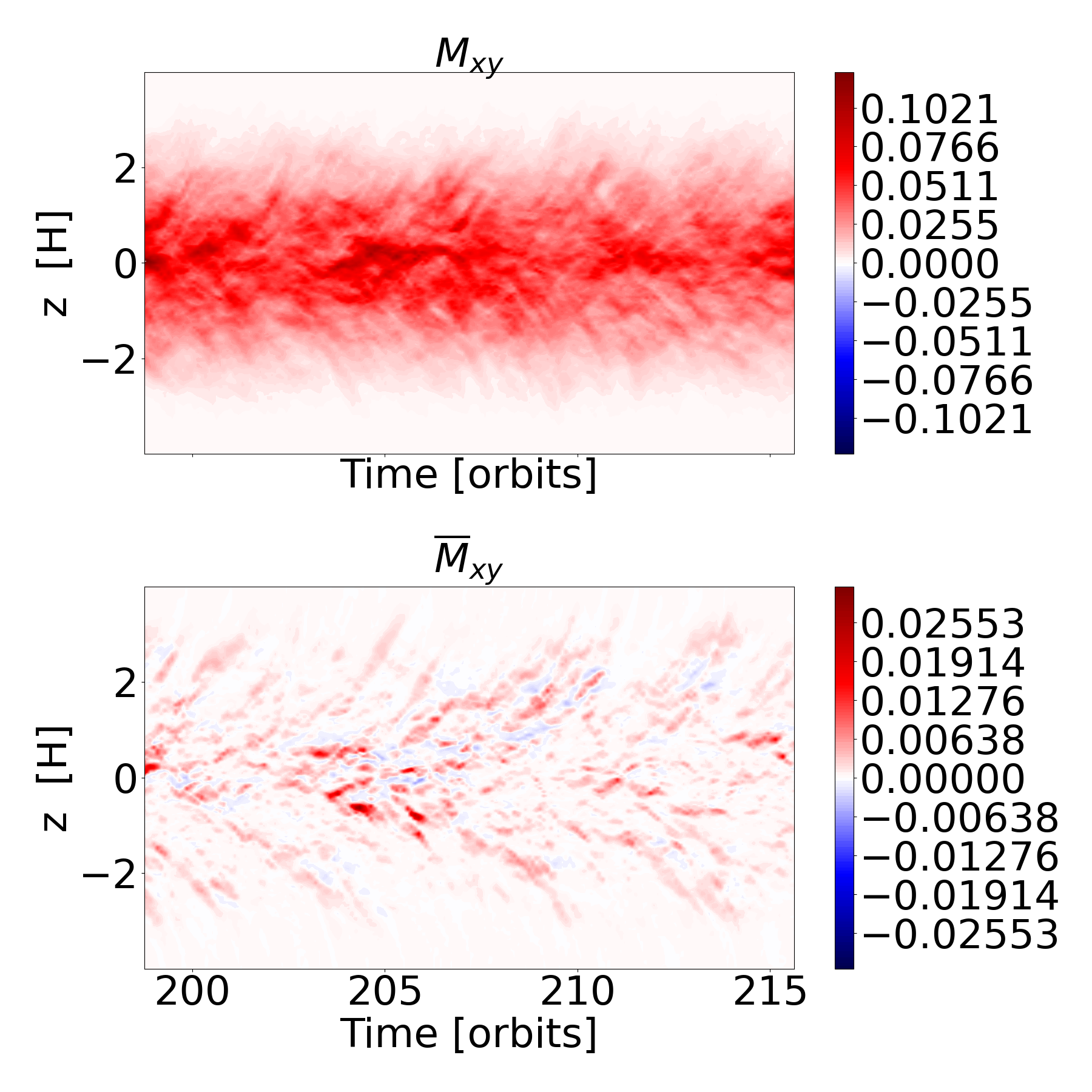}
\caption{Comparison of the butterfly diagrams of the Maxwell stress $M_{xy}$ 
due to the the small-scale magnetic field component (top) and the large-scale (mean) magnetic field component (bottom) for magnetic Prandtl number $\mathrm{Pm}=4$ (left) and $\mathrm{Pm}=90$ (right). The scales for the bottom plots are divided by 4 compared to the top plots. }
\label{fig:stresses}
\end{figure*}

Another important impact of MRI turbulence on disks and stars is the transport of angular momentum. Paper I showed that the angular momentum transport, as parametrized by the ratio of volume-averaged turbulent stress to pressure, increases with $\mathrm{Pm}$. 
In this paper, we would like to look at the region where angular momentum transport is the most important and distinguish the contribution by the small-scale turbulence from the mean-magnetic field. For this, we look at the Maxwell stress $M_{xy}= - B'_x B'_y$ from the small-scales and from the largest scales  $\overline{M}_{xy}= - \overline{B}_x \overline{B}_y$ (Figure \ref{fig:stresses}).
The small-scale angular momentum transport is stronger close to the midplane ($|z|<2H$) in both Pm-regimes, while the large-scale angular momentum transport is more homogeneous in the whole simulation. In terms of amplitudes, the transport from the mean field is $\approx 4$ lower in the midplane so it starts to dominate beyond $|z|>2.5H$.  
This was already shown in the spectral analysis of the Maxwell stress of Paper I  as only the contribution close to $k_x =0$ and $k_y=0$ remained at $z=3.2 H$.
Since the transport from the mean-magnetic field amplitude happens by small patches, its integrated contribution for $\mathrm{Pm}=4$ and $\mathrm{Pm}=90$ contributes respectively to $12\%$-$16\%$ of the total transport, which shows the importance of the mean-field dynamo for angular momentum transport as well.

\section{Comparison with a 3D GRMHD simulation}
\label{sec:Comp}

In this section, we compare the properties of the MRI dynamo in the shearing box simulations to a
3D ideal GRMHD simulation in which the remnant massive neutron star is long-lived \citep{2024Kiuchi}. This simulation uses equatorial symmetry and an $\alpha\Omega$ dynamo was also found in the remnant neutron star to generate a cyclic mean magnetic field, that leads to the launch of a jet. 
In order to compare the two simulations, we take the data from the global simulation averaged azimuthally at $s_0=28.5$ km 
and assume that $H_{GRMHD}= 7$ km so that, in both configurations, the vertical density profile roughly drops by one order of magnitude at the same vertical height. 
Note that the scale height as defined in disks $H=\frac{\Omega}{c_s}$, where $c_s$ is the sound speed, would give a different value $H\approx 15$ km in the neutron star remnant. The shear is also slightly sub-Keplerian ($q=1.34$) in the GRMHD simulations.  

With this value, we compare the correlations above the midplane (Figure \ref{fig:corr_comp}) at $s=28.5$ km and find a good agreement between the shearing box simulations (blue and orange curves) and the GRMHD simulations (green curve). It seems that the global simulation correlations are lower. This could be due to a more turbulent/chaotic dynamo or due to GR effects as lower correlations have been found in different GR simulations \citep{2024Jacquemin-IdeBHGRMHD}. 

\subsection{Dynamo coefficients}

\begin{figure}[ht]
\includegraphics[width=0.45\textwidth]{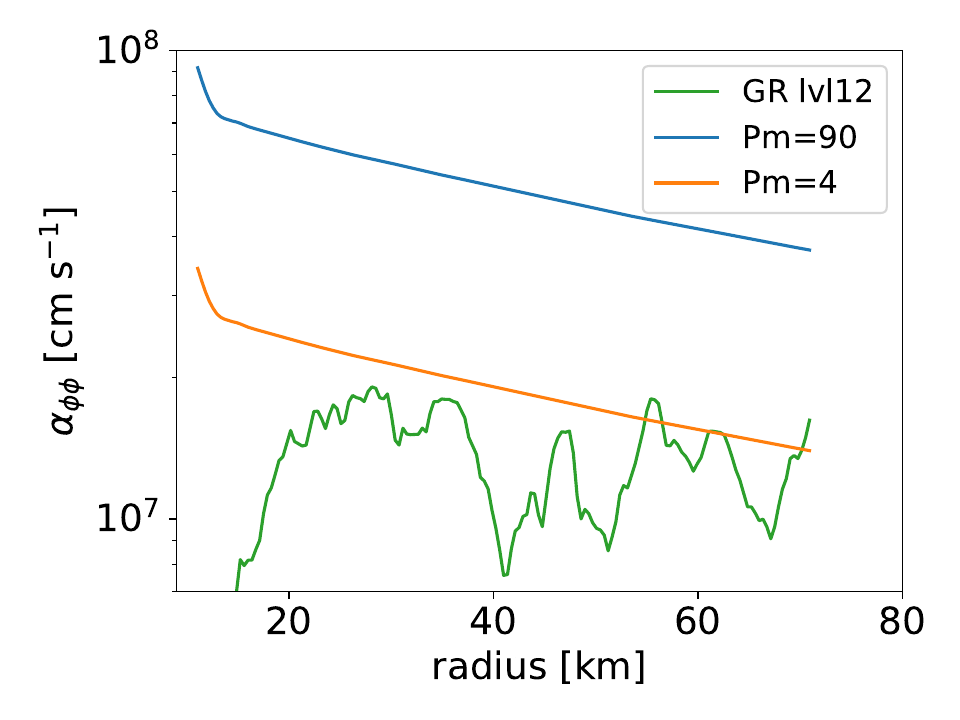}
\caption{Comparison of the maximum value of the diagonal $\alpha_{\phi\phi}$ computed using the IROS method from shearing box simulations at different magnetic Prandtl numbers $\mathrm{Pm}=4$ and 90 (orange and blue curves, respectively) and from the GRMHD simulation (green curve). The sound speed of the GRMHD simulation was used to scale the shearing box simulation results.}
\label{fig:alpha_cs_scaling}
\end{figure}

To have a more quantitative comparison for the dynamo, we use the value of the sound speed averaged between $z\in[0,10]$ km in the binary neutron star simulation to scale the maximum value of the diagonal $\alpha_{yy}$ in shearing box simulations (Figure \ref{fig:alpha_cs_scaling}). The mean-field dynamo in the global simulation has a slightly lower $\alpha$-effect than the $\mathrm{Pm}=4$ shearing box simulation but much lower than the $\mathrm{Pm}=90$. Since we would expect a lower $\alpha$-effect for a shearing box with $\mathrm{Pm}=1$, this shows that the global simulation, which has a numerical Prandtl number of order unity $\mathrm{Pm}_{\rm GRMHD}\approx O(1)$, has a slightly more efficient $\alpha\Omega$ dynamo than the shearing box dynamo.
We also find that the mean-field dynamo in the GRMHD simulations is less efficient at smaller radii where the fluid is denser. This could be due to several reasons. 
First, it is most likely an impact of the shear rate. In the hypermassive neutron star remnant the angular frequency first increases with radius (MRI-stable region) and peaks at $\sim 10$ km, after which it decreases with radius (MRI-unstable region). Thus the shear \textit{rate} ($q \equiv \mathrm{d}\ln{\Omega}/{\mathrm{d}\ln{r}}$) decreases with decreasing radius in the MRI-unstable region and reaches $q=0$ at $\sim 10$ km, which leads to reduced turbulence and dynamo efficiency. 
Finally, GR effects become non-negligible in the core region and might play a role in the decrease in the dynamo efficiency at small radii.
In any case, the comparison shows that if the GRMHD simulation was in the right regime of magnetic Prandtl number of $\mathrm{Pm} \sim 10^{10-14}$ \citep{2017GuiletMergers}, the dynamo period and the mean-field growth rate should be 60\% faster, which would lead to a faster launch of a jet.
In addition, the mean magnetic field should also be further amplified by a factor $\approx 100^{0.744/2} = 5.5$. This could have important astrophysical consequences as the Pointing-flux luminosity and the magnetization parameter of the jet could be stronger by a factor $\sim 30$.

\section{Discussion}
\label{sec:Disc}

\subsection{Disk vs neutron star differences}

The shearing box simulations used in this paper are set up to model MRI turbulence in a disk. However, there are some differences between disks and neutron stars (i.e. protoneutron stars in core-collapse supernovae and hypermassive neutron stars in binary neutron star mergers) that could change the results. The first one is the spherical/cylindrical geometry of PNS/HMNS compared to the cylindrical geometry of disks.
This has been studied by \cite{2021ReboulSalze} who showed that similar results between shearing box and global
simulations of PNS could be obtained in shearing boxes with an azimuthal length $L_y \sim r$, on which the curvature of the star can be neglected. 
The shearing box simulations in this paper here have $L_y=4H$ and therefore match the curvature of a neutron star remnant at a radius $r=4H_{GRMHD} \approx 28$ km. 

Another difference is that the shearing box here neglects the stratification in the radial direction compared to global models, which here would be $r=4H_{GRMHD} \approx 28$ km. On this radial length, the density would change in protoneutron stars or hypermassive neutron stars but it allows to have a bigger domain for the average in the radial direction, which may explain why the correlations are higher than in the global GRMHD simulations.

Lastly, the force balance may be different in protoneutron stars or hypermassive neutron stars than the balance between gravity and centrifugal forces of a disk. This could lead to a lower shear rate than $q=1.5$ and change the scaling of the magnetic field with $\mathrm{Pm}$ \citep[cf.][Figure 12]{2022HeldMRIHighPm}. However, this change in the scaling is found for unstratified boxes with a large aspect ratio. With this configuration, the winding of a strong toroidal field, which highly depends on the shear rates, causes most of the dynamics to sit at large scales and depend less on small scales (i.e. $\mathrm{Pm}$). The dependence of the $\mathrm{Pm}$ scaling on the shear rate might change with cubic boxes or with vertical stratification, which introduces different scales. This is left to future studies.

\subsection{Impact of thermal stratification}

The shearing box simulations here are isothermal. Depending on the timescale of cooling and heating, this is a good approximation. For protoneutron stars or hypermassive neutron stars, the neutrino cooling timescale is of the order O(1-10)s and therefore longer than the dynamo timescale $\sim \frac{2\pi}{\Omega}$ in fast rotating cases. For these objects,  adding thermal stratification could slightly change the results. However, the thermal diffusion due to neutrinos is higher than the viscosity $Pr \equiv \frac{\nu}{\kappa} \approx 10^{-3}-10^{-2}$ \citep{2022ReboulSalze}, which should limit the changes. 
For disks, in order to probe the thermal stability of disks in the large-Pm regime, future work should also take into account the effects of neutrino cooling, thermodynamics and temperature and density-dependent diffusion coefficients.

\section{Conclusions}
\label{sec:Conclusion}

We have carried out a new analysis of shearing box simulations in Paper I in order to investigate the impact of the magnetic Prandtl number $\mathrm{Pm}$ on the MRI-driven mean-field dynamo. We have studied the mean-field dynamo in the regime of high-magnetic Prandtl number, the ratio of viscosity over resistivity, which is relevant for binary neutron star or black hole-neutron star mergers, protoneutron star, and the inner regions of AGN and X-ray binary disks.

Our main result is that the mean-magnetic field follows the scaling law derived for the magnetic energy in \citet{2024HeldMRIPm}. 
We showed that the mean-magnetic field evolution can be described by an $\alpha\Omega$ dynamo, even in the high-Pm regime. We have also computed the dynamo coefficients, which also increase with Pm but with a smaller factor than for the magnetic field. 
We found as well that the dynamo period is also shortened at high Pm and we have a good agreement between the theoretical estimation of the dynamo period and the butterfly diagrams.

We also deduce a different behavior of the mean-field dynamo depending on the vertical height. In the atmosphere, the off-diagonal coefficient $\alpha_{\phi s}$ dominates the contribution to the electromotive force that generates the poloidal magnetic field.
While it does not dominate closer to the mid-plane of the disk,
 this coefficient impacts the propagation of the magnetic field as the turbulent pumping derived from it leads to a faster vertical propagation than the mean vertical velocity. The propagation speed increases with $\mathrm{Pm}$, which can be explained by an increase in the off-diagonal dynamo coefficients.
Overall, we recover the differences between the MRI-driven dynamo and magnetic buoyancies instabilities found in Paper I. 

For systems in the high-Pm regime, the amplification of large-scale magnetic fields happening in ideal MHD, which has a numerical magnetic Prandtl number of order 1, is expected to be realistically further amplified by at least a factor of $5$. 
This increased efficiency of the MRI-driven dynamo could have many important physical consequences: the jet found in GRMHD simulations of binary neutron star mergers could be launched earlier, and be more luminous and more relativistic. In black-hole disk simulations, the magnetically arrested disk state \citep{2011TchekhovskoyMADstateBH} 
could be reached faster, which would facilitate the launch of a relativistic jet. In the context of core-collapse supernovae, the impact of the increased efficiency would be similar: this would more easily lead to greater Lorentz factors in relativistic jets and greater luminosity from the magnetic spin-down of the protoneutron star. 

To study the impact of these enhanced large-scale magnetic fields in global simulations where the high-Pm regime is numerically not reachable, MRI turbulence has to be included using subgrid models. Often, only the angular momentum transport is included using viscous hydrodynamics (alpha-disk models) 
\citep{2013Fernandezalphadisk,2018FujibayashiGRVischydro,2020AFujibayashiGRViscBNS,2020FujibayashiGRViscSN,2023Justalpha}.  
This neglects the other impacts of the mean-magnetic field such as the disk winds, further extraction of rotational energy to power a jet, etc. Some models use a scalar diagonal dynamo coefficient in the mean-field formulation to have the amplification and the dynamo cycle \citep{2022DelZannaGRMHDMeanField}, but the dynamo coefficients or the saturated magnetic energy are often lower than what is found in this paper \citep{2021ShibataGRMHDalphaOmega,2023MostMeanField}. Moreover, the astrophysical consequences of the addition of off-diagonal dynamo coefficients remain unknown and could help get closer results to 3D global simulations.

\begin{acknowledgements}
 Simulations were run on the Sakura, Cobra, and Raven clusters at the Max Planck Computing and Data Facility (MPCDF) in Garching, Germany. 
 This work was in part supported by the Grant-in-Aid for Scientific Research (grant Nos. 23H04900, 23H01172, 23K25869) of Japan MEXT/JSPS. 
\end{acknowledgements}

%
%
   \bibliographystyle{aa} 
   \bibliography{biblio} 

\end{document}